\documentclass{article}
\usepackage{ijcai21}

\usepackage{times}
\usepackage{soul}
\usepackage{url}
\usepackage[hidelinks]{hyperref}
\usepackage[utf8]{inputenc}
\usepackage[small]{caption}
\usepackage{graphicx}
\usepackage{amsmath}
\usepackage{amsthm}
\usepackage{booktabs}
\usepackage{algorithm}
\usepackage{algorithmic}
\usepackage{booktabs}   
\usepackage{relsize}
\usepackage{adjustbox}
\usepackage{tikz}
\usepackage{pgfplots}
\usepackage{hyperref}
\usepackage{flushend} 
\usepackage{caption}
\usepackage{alltt}
\usepackage{csquotes}
\floatstyle{ruled}
\newfloat{program}{t}{lop}
\floatname{program}{Program}
\usepackage{relsize}
\usepackage{adjustbox}
\usepackage{tikz}
\usepackage{pgfplots}
\usepackage{multicol}
\usepackage{calc}
\usepackage{ifthen}
\usepackage{datapie}
\newcommand{\etal}{\textit{et al.}}
\newcommand{\smartinduct}{\texttt{smart\_induct}}
\newcommand{\smarterinduct}{\texttt{sem\_ind}}
\newcommand{\induct}{the \texttt{induct} tactic}

\newcommand{\selfie}{SeLFiE}

\pgfplotsset{compat=1.16}
\usepackage{soul}
\usepackage{tikz}
\usepackage{pgfplots}
\usepackage{calc}

\title{Faster Smarter Proof by Induction in Isabelle/HOL}

\author{
Yutaka Nagashima\\
\affiliations
Yale-NUS College, National University of Singapore\\
University of Innsbruck\\
Czech Institute of Informatics, Robotics, and Cybernetics, Czech Technical University in Prague\\
\emails
yutaka@yale-nus.edu.sg
}

\begin{document}

\maketitle

\begin{abstract}
We present \smarterinduct{},
a recommendation tool for proof by induction in Isabelle/HOL.
Given an inductive problem,
\smarterinduct{} produces candidate arguments for proof by induction,
and selects promising ones using heuristics.
Our evaluation based on 1,095 inductive problems from 22 source files shows that
\smarterinduct{} improves the accuracy of recommendation from
20.1\% to 38.2\%
for the most promising candidates
within 5.0 seconds of timeout compared to its predecessor
while decreasing the median value of execution time
from 2.79 seconds to 1.06 seconds.
\end{abstract}

\section{Introduction}\label{sec:intro}

As our society grew reliant on software systems,
the trustworthiness of such systems became essential.
One approach to develop trustworthy systems is complete formal verification using proof assistants.
In a complete formal verification,
we specify the desired properties of our systems and
prove that our implementations are correct in terms of the specifications using
software tools, called \textit{proof assistants}. 

In many verification projects,
proof by induction plays a critical role.
To facilitate proof by induction,
modern proof assistants offer sub-tools, called \textit{tactic}s.
For example,
Isabelle \cite{isabelle} comes with \induct{}.
Using \induct{},
human proof authors can apply proof by induction
simply by passing appropriate arguments
instead of manually developing induction principles.
When choosing such arguments,
proof engineers have to answer the following three questions:

\begin{itemize}
  \item On which terms do they apply induction?
  \item Which variables do they pass to the \texttt{arbitrary} field to generalise them?
  \item Which induction rule do they pass to the \texttt{rule} field?
\end{itemize}

For example, 
Program \ref{p:motivating_example}
defines 
the append function (\verb|@|) and
two reverse functions (\verb|rev1| and \verb|rev2|)
and 
presents two ways to 
prove their equivalence
by applying \induct{}.
Note that
\verb|[]|, \verb|#|, and \texttt{[x]}
represent the empty list, the list constructor, and
the syntactic sugar for \texttt{x \# []}, respectively.

\begin{program}
\begin{alltt}
@ :: \(\alpha\) list \(\Rightarrow\) \(\alpha\) list \(\Rightarrow\) \(\alpha\) list
  []       @ ys = ys
| (x # xs) @ ys = x # (xs @ ys)

rev1 :: \(\alpha\) list \(\Rightarrow\) \(\alpha\) list 
  rev1 []       = []
| rev1 (x # xs) = rev1 xs @ [x]

rev2 :: \(\alpha\) list \(\Rightarrow\) \(\alpha\) list \(\Rightarrow\) \(\alpha\) list
  rev2 []       ys = ys
| rev2 (x # xs) ys = rev2 xs (x # ys)

theorem rev2 xs ys = rev1 xs @ ys
 apply(induct xs ys rule: rev2.induct)
by auto

theorem rev2 xs ys = rev1 xs @ ys
 apply(induct xs arbitrary: ys) by auto
\end{alltt}

\caption{Equivalence of two reverse functions}
\label{p:motivating_example}
\end{program}

The first proof script applies computation induction
by passing \texttt{rev2.induct} to the \texttt{rule} field.
\texttt{rev2.induct} is a customised induction rule,
which Isabelle automatically derives from the definition of \texttt{rev2}.
The subsequent application of \texttt{auto} discharges all sub-goals produced 
by this induction.

The second proof script applies structural induction on \verb|xs| 
while generalising \verb|ys|.
This application of structural induction results in the following base case and induction step:

\begin{alltt}
base case: rev2 [] ys = rev1 [] @ ys
induction step:
 (\(\forall\)ys. rev2 xs ys = rev1 xs @ ys) \(\longrightarrow\)
 rev2 (a # xs) ys = rev1 (a # xs) @ ys
\end{alltt}
\noindent
where $\forall$ and $\longrightarrow$ represent the universal quantifier and implication, respectively.
Using the associative property of \texttt{@},
the subsequent application of 
\texttt{auto} firstly transformed the induction step to the following intermediate goal internally:
\begin{alltt}
(\(\forall\)ys. rev2 xs ys = rev1 xs @ ys) \(\longrightarrow\) 
rev2 xs (a # ys) = rev1 xs @ (a # ys)
\end{alltt}

\noindent
Since \texttt{ys} was generalised in the induction hypothesis,
\texttt{auto} proved \texttt{rev2 xs (a} \verb|#| \texttt{ys) = rev1 (xs} \verb|@| \texttt{(a} \verb|#| \texttt{ys))}
by considering it as a concrete case of the induction hypothesis.
If we remove \texttt{ys} from the \texttt{arbitrary} field,
the subsequent application of \texttt{auto} leaves the induction step as follows:
\begin{alltt}
rev2 xs ys = rev1 xs @ ys \(\longrightarrow\) 
rev2 xs (a # ys) = rev1 xs @ (a # ys)
\end{alltt}

In other words,
\texttt{auto} cannot make use of the induction hypothesis
since the conclusion of induction step
share the \textit{same} \texttt{ys}.
Experienced human researchers can judge that
this application of \induct{} was not appropriate.
However, it is also true that
this induction step is still provable.
For this reason, counter-example finders, such as Nitpick \cite{nitpick} and Quickcheck \cite{quickcheck},
cannot detect that
this \texttt{induct} tactic without generalisation is not appropriate
for this problem.
This is why
engineers still have to carefully examine inductive problems to 
answer the aforementioned three questions
when using \induct{}.

This issue is not specific to Isabelle:
other proof assistants, such as Coq \cite{coq}, HOL4 \cite{hol4}, and HOL Light \cite{hollight},
offer similar tactics for inductive theorem proving,
and it is human engineers who have to specify the arguments for such tactics.
This issue is not trivial either:
in a summary paper from 2005,
Gramlich listed generalisation as one of \textit{the main problems and challenges}
of inductive theorem proving 
while predicting that 
\textit{
substantial progress in inductive theorem proving will take time}
due to 
\textit{the enormous problems and the inherent difficulty of inductive theorem proving}
\cite{gramlich}.

Previously, we built \smartinduct{},
which suggests arguments of \induct{}
in Isabelle/HOL.
Our evaluation 
showed that \smartinduct{} predicts
on which variables Isabelle experts apply \induct{} for some inductive problems.
Unfortunately, \smartinduct{} has the following limitations:

\begin{itemize}
    \item[L1.] It tends to take too long to produce recommendations.
    \item[L2.] It cannot recommend induction on compound terms or induction on multiple occurrences of the same variable.
    \item[L3.] It is bad at predicting variable generalisations.
    \item[L4.] Its evaluation is based on a small dataset with 109 inductive problems.
\end{itemize}

We overcame these problems with \smarterinduct{},
a new recommendation tool for \induct{}.
Similarly to \smartinduct{},
\smarterinduct{} suggests what arguments to pass to the \induct{}
for a given inductive problem.
Our overall contribution is that 
\begin{displayquote}
we built a system
that predicts how one should apply proof by induction
in Isabelle/HOL both quickly and accurately.
\end{displayquote}

Even though we built \smarterinduct{} for Isabelle/HOL,
our approach is transferable to other proof assistants
based on tactics:
no matter what proof assistants we use,
we need an architecture 
that aggressively removes less promising candidates
to address L1 (presented in Section \ref{sec:architecture}),
a procedure to construct promising induction candidates
without missing out too many good ones
to address L2
(presented in Section \ref{sec:smart_construct}),
and
domain-agnostic heuristics
that analyse not only the syntactic structures of
inductive problems but the definitions of relevant constants
to address L3
(presented in Section \ref{sec:heuristic}).
Finally, Section \ref{sec:evaluation} justifies our claims through extensive evaluations 
based on 1095 inductive problems, addressing L4.

\section{The Overall Architecture}\label{sec:architecture}

\begin{figure}[t]
      \centerline{\includegraphics[width=1.0\linewidth]{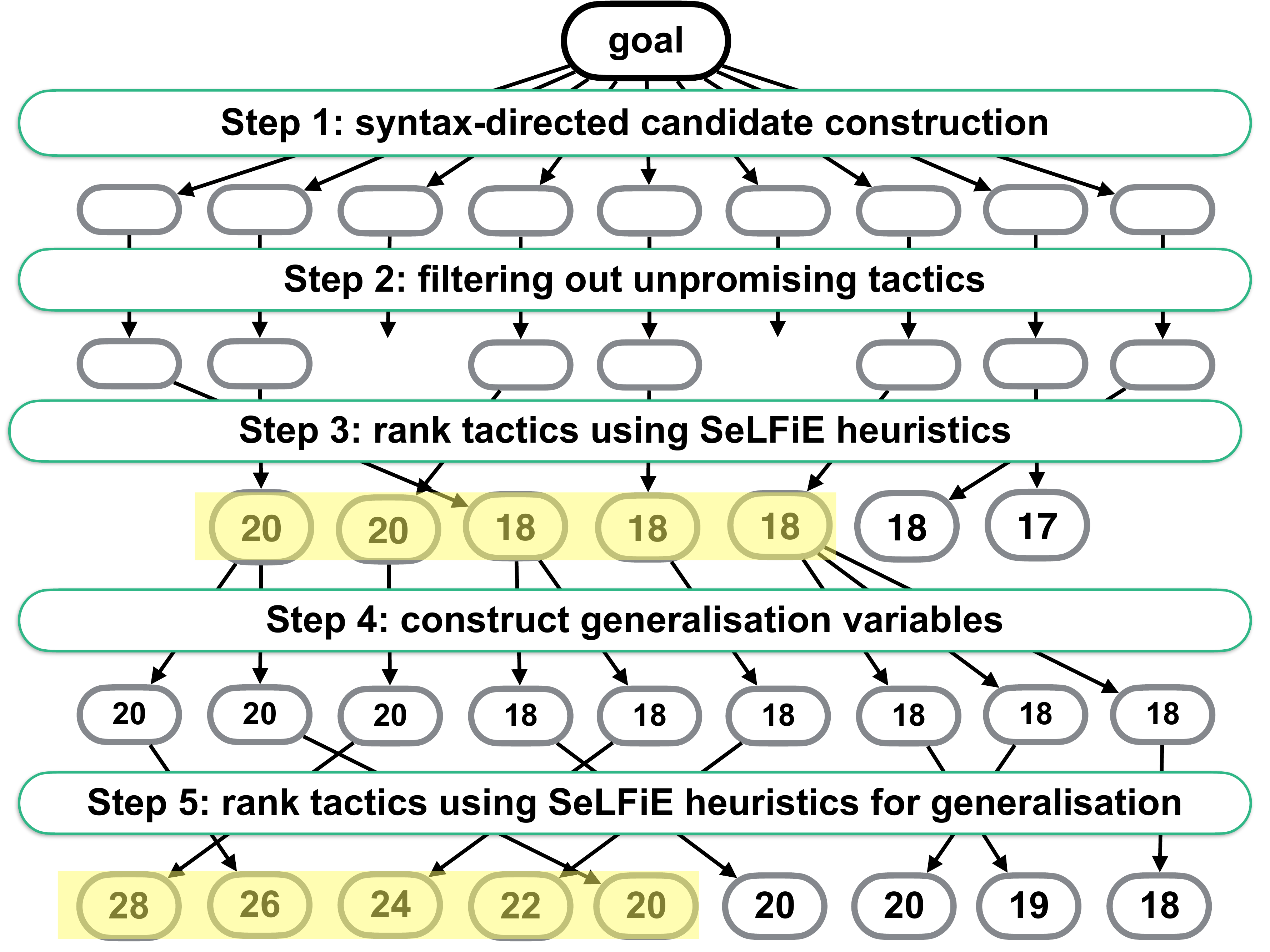}}
      \caption{The overview of \smarterinduct{}.}
      \label{fig:semantic_induct}
\end{figure}

Figure \ref{fig:semantic_induct} illustrates the overall architecture of \smarterinduct{}, consisting of 5 steps to
produce and select candidate tactics as follows.

\paragraph{Step 1.}
\smarterinduct{} produces a set of sequences of induction terms and induction rules for \induct{}
    from a given inductive problem.
The aim of this step is to produce a small number of candidates intelligently,
so that it covers most promising sequences of induction terms and induction rules
while avoiding a combinatorial blowup.
We expound the algorithm to achieve this goal in Section \ref{sec:smart_construct}.

\paragraph{Step 2.} 
\smarterinduct{} applies \induct{} with the sequences of arguments produced in Step 1 to discard less promising candidates.
\smarterinduct{} decides a sequence of arguments is unpromising 
if the sequence satisfies any of the following conditions:
\begin{itemize}
    \item \induct{} fails with an error message, or
    \item one of the resulting intermediate goals is identical to the original goal itself.
\end{itemize}
    
\paragraph{Step 3.} 
\smarterinduct{} applies 36 pre-defined heuristics written in \selfie{} \cite{selfie},
which we explain in Section \ref{sec:heuristic} using our running example.
These heuristics judge the validity
of induction terms and induction rules with respect to the proof goal and
the relevant definitions.
Each heuristic is implemented as an assertion
on inductive problems and arguments of \induct{},
and each assertion is tagged with a value.
If an assertion returns \verb|True| for a sequence of arguments, 
\smarterinduct{} gives the tagged value to the sequence.
\smarterinduct{} sums up such points from the 36 heuristics
to compute the score for each sequence.
Based on these scores, \smarterinduct{} sorts sequences of arguments from Step 2
and selects the five most promising sequences for further processing.

\paragraph{Step 4.}
After deciding induction terms and induction rules for \induct{} in Step 3,
\smarterinduct{} adds arguments for the arbitrary field to the sequences
of arguments passed from Step 3.
Firstly,
\smarterinduct{} collects free variables in the proof goal 
that are not induction terms for each sequence from Step 3.
Then, it constructs the powerset of such free variables and
uses each set in the powerset as the arguments to the \texttt{arbitrary} field of \induct{}.
For example, if \smarterinduct{} receives \texttt{(induct xs)} from Step 3
for our running example of list reversal,
it produces \verb|{|\verb|{|\verb|}, |\verb|{|\texttt{ys}\verb|}|\verb|}|
as the powerset 
because \texttt{xs} and \texttt{ys} are the only free variables in the goal
and \texttt{xs} appears as the induction term.
This powerset leads to the following two \texttt{induct} tactics: \texttt{(induct xs)}, and \texttt{(induct xs arbitrary:ys)}.

\paragraph{Step 5.}
For each remaining sequence, \smarterinduct{} applies 
8 pre-defined \selfie{} heuristics to judge the validity
of generalisation.
Again, each heuristic is tagged with a value,
which is used to compute the final score for each candidate:
for each sequence, 
\smarterinduct{} adds the score from the generalisation heuristics
to the score from Step 3 to decide the final score for each sequence.
Based on these final scores, 
\smarterinduct{} sorts sequences of arguments from Step 5
and presents the 10 most promising sequences in the Output panel of Isabelle/jEdit,
the default proof editor for Isabelle/HOL \cite{jedit}.

\begin{figure}[t]
      \centerline{\includegraphics[width=0.9\linewidth]{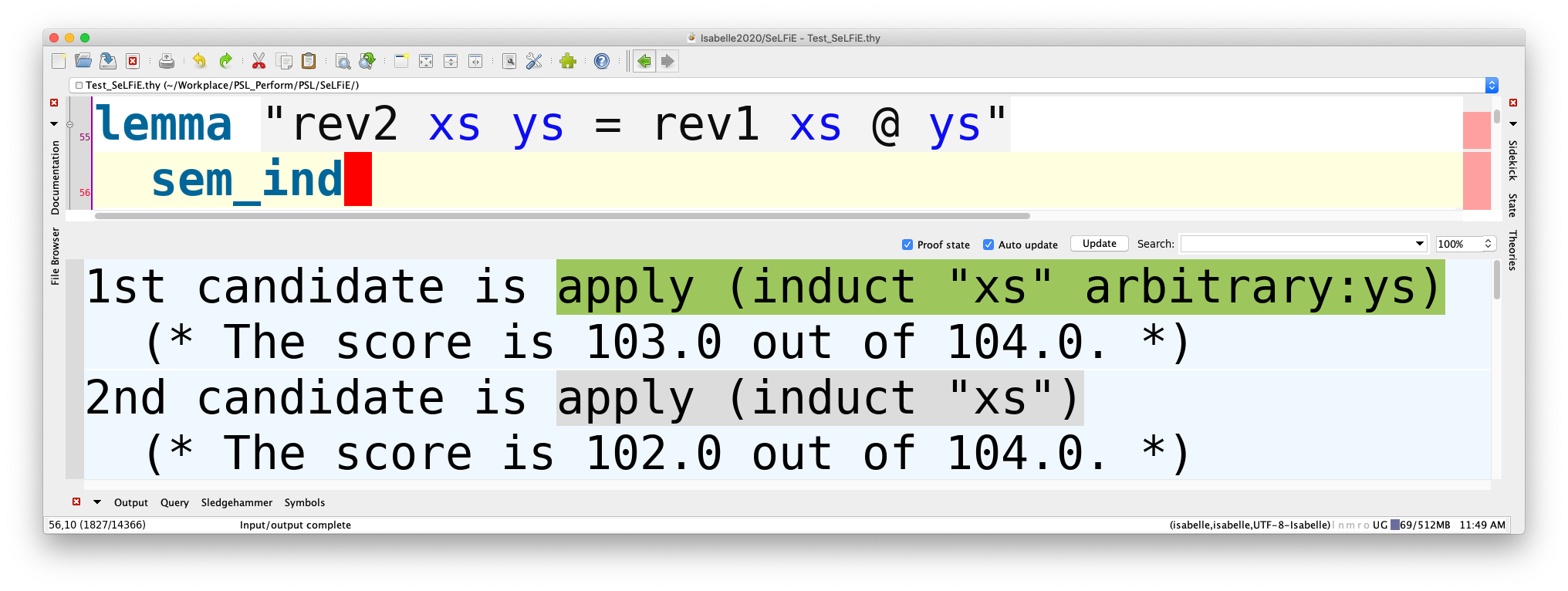}}
      \caption{The user-interface of \smarterinduct{}.}
      \label{fig:screenshot}
\end{figure}

We developed \smarterinduct{} entirely 
within the Isabelle ecosystem without any dependency to external tools.
This allows for the easy installation process of \smarterinduct{}:
to use \smarterinduct{}, users only have to download the relevant Isabelle files from our public GitHub repository\footnote{\url{https://github.com/data61/PSL}}
and install \smarterinduct{} using the standard Isabelle command.
The seamless integration into the Isabelle proof language, Isar \cite{isar}, 
lets users invoke \smarterinduct{} within their ongoing proof development
and copy a recommended \texttt{induct} tactic to the right location
with one click as shown in Figure\ \ref{fig:screenshot}.

\section{Syntax-Directed Candidate Construction}\label{sec:smart_construct}

In general, \induct{} may take multiple induction terms and induction rules in one invocation.
However, it is rarely necessary to pass multiple induction rules to the \texttt{rule} filed.
Therefore, \smarterinduct{} passes up-to-one induction rule to \induct{}.
    
On the other hand, it is often necessary to pass multiple induction terms to \induct{},
and the order of such induction terms is important to apply \induct{} effectively.
Moreover, it is sometimes indispensable to pass the same induction term
multiple times to \induct{}, so that each of them corresponds to
a distinct occurrence of the same term in the proof goal.
What is worse, induction terms do not have to be variables: 
they can be compound terms such as function applications.

Enumerating all possible sequences of induction terms leads to a combinatorial explosion.
To avoid such combinatorial explosion, \smarterinduct{} produces
sequences of induction terms and induction rules taking
a syntax-directed approach, which traverses the syntax tree representing the proof
goal while collecting plausible sequences of induction terms and rules
as follows.

\paragraph{Step 1-A.}
The collection starts at the root node of the syntax tree
with an empty set of sequences of induction arguments.

\paragraph{Step 1-B.}
If the current node is a function application,
\smarterinduct{} takes arguments to the function,
produces a set of lists of such arguments while preserving their order.
This set of lists represents candidates for induction terms.
If the function in this function application is a constant with
a relevant induction rule stored in the proof context,
\smarterinduct{} produces candidate \texttt{induct} tactics
with and without this rule for the \texttt{rule} field.

For example, if the current node is \texttt{rev2 xs ys},
Step 1-B produces
\texttt{(induct xs)} and \texttt{(induct xs ys rule:rev2.induct)},
as well as other candidates such as
\texttt{(induct xs ys)} and \texttt{(induct xs rule:rev2.induct)}.

\paragraph{Step 1-C.}
If any sub-terms of the current node is a compound term,
\smarterinduct{} moves down to such sub-terms in the syntax tree
and repeats S1-b to collect more candidates for induction arguments.

\paragraph{Step 1-D.}
\smarterinduct{} finishes Step 1 when it reaches the leaf nodes
in all branches of the syntax tree.

\paragraph{}
    This syntax-directed argument construction avoids a combinatorial explosion
    at the cost of missing out some effective sequences of induction arguments.
    One notable example is the omission of simultaneous induction,
    which is essential to tackle inductive problems with mutually recursive
    functions.
    Our evaluation results in Section \ref{sec:evaluation} show that
    despite the omission of such cases
    \smarterinduct{} manages to recommend correct induction arguments
    for most of the cases that appear in day-to-day theorem proving.
    
    In principle, this smart construction of candidate \texttt{induct} tactics
    ignores handcrafted induction rules:
    in Step 1-B \smarterinduct{} collects induction rules derived automatically
    by Isabelle
    when defining relevant functions.
    For example, \smarterinduct{} picks up \texttt{rev2.induct}
    when seeing \texttt{rev2} in our running example
    since \texttt{rev2.induct} is an induction principle
    Isabelle automatically derived
    when defining \texttt{rev2}.
    This constraint is unavoidable since we cannot predict 
    what induction rules proof engineers will manually develop in the future
    for problem domains that may not even exist yet.
    
    The notable exceptions to this principle are induction rules manually developed in
    Isabelle's standard library:
    in Step 1-B the smart construction algorithm 
    collects some manually developed induction rules from the standard library
    if the rules seem to be relevant to the inductive problem at hand.
    This optimisation is reasonable: 
    some concepts in the standard library, 
    such as lists and natural numbers, are used in many projects and have 
    useful induction rules handcrafted by Isabelle experts.

\section{Induction and Generalisation Heuristics}\label{sec:heuristic}

We now have a closer look at heuristics used in Step 3 and Step 5.
To produce accurate recommendations quickly,
heuristics for \smarterinduct{} have to satisfy the following two criteria.
\begin{itemize}
    \item [C1:] The heuristics should be applicable to a wide range of problem domains,
    some of which do not exist yet.
    \item [C2:] They should be able to analyse not only the syntactic structures
    of the inductive problems at hand but also the definitions of relevant constants
    in terms of how such constants are used within the inductive problems.
\end{itemize}

\begin{figure}[t]
      \centerline{\includegraphics[width=1.0\linewidth]{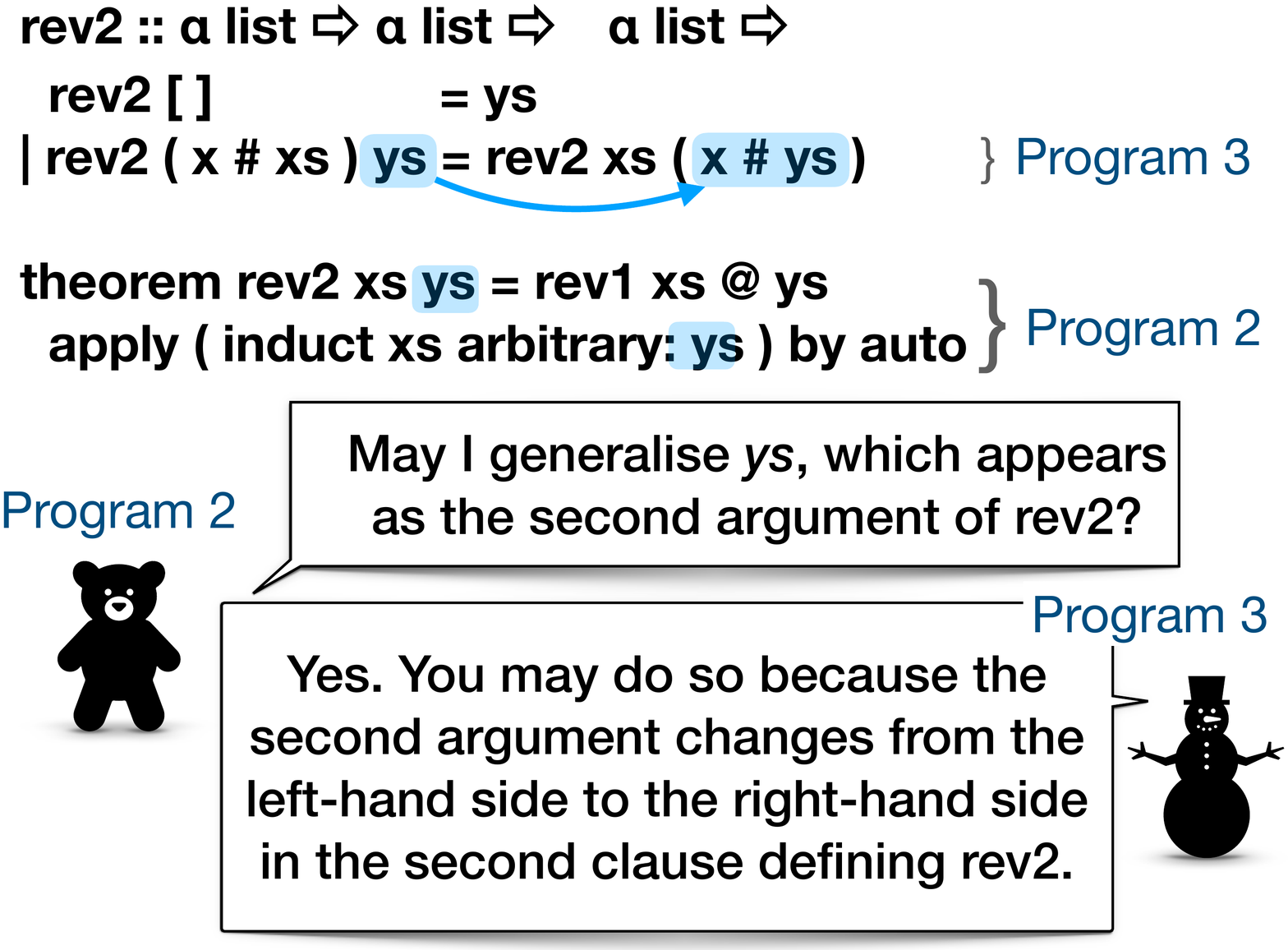}}
      \caption{Semantic-aware generalisation heuristic as a dialogue.}
      \label{fig:dialogue}
\end{figure}

\begin{program}
\begin{alltt}
\(\forall\) \(arb_term\) : term \(\in\) arbitrary_term.
 \(\exists\) \(f_term\) : term.
  \(\exists\) \(f_occ\) : term_occ \(\in\) \(f_term\).
   \(\exists\) \(arb_occ\) \(\in\) \(arb_term\).
    \(\exists\) \(generalise_nth\) : number.
      is_or_below_nth_argument_of
        (\(arb_occ\), \(generalise_nth\), \(f_occ\))
     \(\land\)
      \(\exists{}\sb{def}\)
        (\(f_term\), 
         generalise_nth_argument_of,
         [\(generalise_nth\), \(f_term\)])
\end{alltt}

\caption{Syntactic analysis for generalisation in \selfie{}}
\label{p:generalisation_heuristic_in_selfie_outer}
\end{program}

\begin{program}
\begin{alltt}
generalise_nth_argument_of :=
\(\lambda\) [\(generalise_nth\), \(f_term\)].
 \(\exists\) \(lhs_occ\) : term_occ.
  is_left_hand_side (\(lhs_occ\))
 \(\land\)
  \(\exists\) \(nth_param_on_lhs\) : term_occ.
    is_nth_argument_of
     (\(nth_param_on_lhs\), \(generalise_nth\),
      \(lhs_occ\))
   \(\land\)
    \(\exists\) \(nth_param_on_rhs\) : term_occ.
      \(\neg\) are_of_same_term 
       (\(nth_param_on_rhs\), \(nth_param_on_lhs\))
     \(\land\)
      \(\exists\) \(f_occ_on_rhs\) : term_occ \(\in\) \(f_term\).
       is_nth_argument_of 
         (\(nth_param_on_rhs\),
          \(generalise_nth\),
          \(f_occ_on_rhs\))
\end{alltt}
\caption{Semantic analysis for generalisation in \selfie{}}
\label{p:generalisation_heuristic_in_selfie_inner}
\end{program}

To satisfy the above criteria,
we choose \selfie{} \cite{selfie} as 
our implementation language to encode heuristics.
\selfie{} is is a meta-language to encode heuristics for inductive theorem proving
as assertions.
A \selfie{} assertion takes a pair of arguments to \induct{} and an inductive problem with relevant definitions.
The assertion should return \texttt{True}
if the choice of argument of \induct{} is compatible with the heuristic.

The exact definitions of our 44 heuristics are not informative or possible due to the page limit.
Therefore, instead of presenting each heuristic,
this section introduces one simple generalisation heuristic written in \selfie{}
to demonstrate how we address the above two criteria using \selfie{}.

Program \ref{p:generalisation_heuristic_in_selfie_outer} 
and \ref{p:generalisation_heuristic_in_selfie_inner} 
define the following generalisation heuristic introduced by Nipkow \etal \cite{isabelle}\footnote{
In this explanation we simplified the heuristic to focus on the essence of \selfie{}.
The corresponding heuristic we used for \smarterinduct{} involves
optimisations and handling of corner cases.}:

\begin{displayquote}
(Variable generalisation) should not be applied blindly. 
It is not always required, 
and the additional quantifiers can complicate matters in some cases. 
The variables that need to be quantified are typically those that change in recursive calls.
\end{displayquote}

Figure \ref{fig:dialogue} illustrates how this heuristic
justifies the generalisation of \texttt{ys} 
in our running example as an informal dialogue between 
Program \ref{p:generalisation_heuristic_in_selfie_outer} and Program \ref{p:generalisation_heuristic_in_selfie_inner}.
In this dialogue, Program \ref{p:generalisation_heuristic_in_selfie_outer}
analyses the syntactic structure of the proof goal in terms of the arguments of \induct{},
whereas Program \ref{p:generalisation_heuristic_in_selfie_inner} analyses the
definition of \texttt{rev2} in terms of how \texttt{rev2} is used in the goal.
Note that Program \ref{p:generalisation_heuristic_in_selfie_outer} and
Program \ref{p:generalisation_heuristic_in_selfie_inner} realise this dialogue
through a \textit{definitional quantifier}, $\exists{}_{def}$.
With this dialogue in mind, we now formally interpret
the two programs for our running example.

Program \ref{p:generalisation_heuristic_in_selfie_outer} checks
for all generalised variable, \textit{arb\_term},
if there exists a function, \textit{f\_term}, 
its occurrence, \textit{f\_occ},
an occurrence of the generalised variable, \textit{arb\_occ}, 
and a natural number, \textit{generalise\_nth},
that satisfy the conjunction.
Since our running example has only one generalised variable, \texttt{ys}, 
if we choose
\begin{itemize}
    \item \texttt{rev2} for \textit{f\_term},
    \item the only occurrence of \texttt{rev2} in the proof goal for \textit{f\_occ},
    \item the occurrence of \texttt{ys} on the left-hand side of the equation in the proof goal for \textit{arb\_occ},
    \item $2$ for \textit{generalise\_nth},
\end{itemize}
we can satisfy the first conjunct for all \textit{arb\_term}s
because in the proof goal \textit{ys} appears as the second argument
of the only occurrence of \texttt{rev2}.

In the second conjunct, 
Program \ref{p:generalisation_heuristic_in_selfie_outer}
uses $\exists{}_{def}$
to ask Program \ref{p:generalisation_heuristic_in_selfie_inner}
if there exists a clause defining \texttt{rev2}
that satisfies the condition specified in Program \ref{p:generalisation_heuristic_in_selfie_inner}.
Formally speaking,
Program \ref{p:generalisation_heuristic_in_selfie_inner} examines
if there is a term occurrence, \textit{nth\_param\_on\_lhs},
such that
\textit{nth\_param\_on\_lhs} is the \textit{generalise\_nth} argument on the left-hand side
in the equation defining \textit{f\_term}, 
but an occurrence of \textit{f\_term} on the right-hand side has a different
term for its second argument.

In the second clause defining \texttt{rev2},
the second argument of \texttt{rev2}
is \texttt{ys} on the left-hand side, 
while the second argument of \texttt{rev2} is \texttt{x\#ys} on the right-hand side.
Therefore, Program \ref{p:generalisation_heuristic_in_selfie_inner}
returns \texttt{True} to
$\exists{}_{def}$ in Program \ref{p:generalisation_heuristic_in_selfie_outer},
with which Program \ref{p:generalisation_heuristic_in_selfie_outer}
confirms that the candidate arguments of \induct{} satisfy Nipkow's heuristic.

Attentive readers may have noticed that
Program \ref{p:generalisation_heuristic_in_selfie_outer}
and Program \ref{p:generalisation_heuristic_in_selfie_inner}
satisfy the aforementioned two criteria.
They satisfy C1 because
they refer to problem specific constants and arguments, such as \texttt{rev2} and \texttt{ys},
abstractly using quantifiers,
so that they can be applicable to other inductive problems.
They also satisfy C2 because
Program \ref{p:generalisation_heuristic_in_selfie_inner}
analyses the definitions of relevant constants, such as \texttt{rev2},
while Program \ref{p:generalisation_heuristic_in_selfie_outer}
analyses the syntactic structures of the problem.
This is why \smarterinduct{} achieves higher coincidence rates
compared to its predecessor, \smartinduct{} \cite{smart_induction},
reported in Section \ref{sec:evaluation}.

In total, we implemented 44 heuristics in \selfie{}.
36 of them are induction heuristics and 8 of them are generalisation heuristics.
We adopted some heuristics from \smartinduct{},
and we newly implemented others based on literature and our expertise.
As discussed above,
\selfie{} allows us to encode heuristics that transcend problem domains
using quantifiers.
At the same time, however, 
some basic concepts, such as lists, sets and natural numbers,
appear in a wide range of verification projects.
Therefore, we developed 20 \selfie{} heuristics 
that explicitly refer to 
concrete constants or manually derived induction rules from the standard library
to improve the accuracy of recommendations for problems involving
such commonly used concepts.
Unlike other parts of this paper,
these 20 heuristics involve expertise specific to Isabelle/HOL.

\section{Evaluation}\label{sec:evaluation}

We evaluated \smarterinduct{} against \smartinduct{}.
Our focus is to measure the accuracy of recommendations and
execution time necessary to produce recommendations.
All evaluations are conducted on a MacBook Pro (15-inch, 2019)
with 2.6 GHz Intel Core i7 6-core memory 32 GB 2400 MHz DDR4.

Unfortunately, it is, in general, not possible to decide
whether a given application of \induct{} is right for a given problem.
In particular, even if we can finish a proof search after applying
\induct{}, this does not guarantee that the arguments passed to \induct{} are a good choice.
For example, it is possible to prove our motivating example by applying
\texttt{(induct ys)};
however, the necessary proof script following this application of \induct{} becomes
unnecessarily lengthy.

Therefore, we adopt \textit{coincidence rates} as the surrogate for
success rates
to approximate the accuracy of \smarterinduct{}'s recommendations:
we measure how often recommendations of \smarterinduct{} \textit{coincide}
with the choice of human engineers.
Since there are often multiple equally valid sequences of induction arguments 
for a given inductive problem,
we should regard coincidence rates as conservative estimates of
true success rates.

As our evaluation target,
we use 22 Isabelle theory files with 1,095 applications of \induct{}
from the Archive of Formal Proofs (AFP) \cite{AFP}.
The AFP is an online repository of formal proofs in Isabelle/HOL.
Each entry in the AFP is peer-reviewed by Isabelle experts prior to acceptance,
which ensures the quality of our target theory files.
Therefore, 
if \smarterinduct{} achieves higher coincidence rates for our target theory files,
we can say that \smarterinduct{} produces good recommendations for many problems.
To the best of our knowledge,
this is the most diverse dataset used to measure recommendation tools for 
proof by induction.
For example, when Nagashima evaluated \smartinduct{}
they used only 109 invocations of \induct{} from 5 theory files, 
all of which are included in our dataset.

\subsection{Coincidence Rates within 5.0 Seconds}\label{sec:eval_result_coincidence}

\begin{table}
\begin{tabular}{c | r r r r}
\hline
\noalign{\smallskip}
tool & top 1 & top 3 & top 5 & top 10\\
\hline
\smarterinduct{}  & 38.2\% & 59.3\% & 64.5\% & 72.7\%\\ 
\smartinduct{}    & 20.1\% & 42.8\% & 48.5\% & 55.3\%\\ 
\hline
\end{tabular}
\caption{Overall coincidence rates within 5.0 seconds of timeout.}\label{table:semantic_induct_coincidence}
\end{table}

Table \ref{table:semantic_induct_coincidence} shows 
coincidence rates of
both \smarterinduct{} and \smartinduct{}
within 5.0 seconds of timeout.

For example, the coincidence rate of \smarterinduct{} is 38.2\% for \textit{top 1}.
This means that the sequences of induction arguments used by human researchers 
appear as the most promising sequences recommended by \smarterinduct{}
for 38.2\% of the uses of \induct{}.
On the other hand, the coincidence rate of \smartinduct{} is 20.1\% for \textit{top 1}.
This means that \smarterinduct{} achieved a 90.0\% increase of the coincidence rate
for the most promising candidates.
Overall, Table \ref{table:semantic_induct_coincidence} indicates that
\smarterinduct{} consistently outperforms \smartinduct{}
when they can suggest multiple sequences.

\begin{table}
\begin{tabular}{c | r r r r}
\hline
\noalign{\smallskip}
tool & top 1 & top 3 & top 5 & top 10\\
\hline
\smarterinduct{}  &  54.5\% & 63.6\% & 72.7\% & 72.7\%\\
\smartinduct{}    &   0.0\% &  0.0\% &  0.0\% &  9.1\%\\ 
\hline
\end{tabular}
\caption{Coincidence rates 
for \texttt{Nearest\_Neighbors.thy}.}\label{table:nearest}
\end{table}

We leave the coincidence rates for each theory file in the accompanying technical appendix \cite{appendix}
but present coincidence rates for a representative theory file in Table \ref{table:nearest}.
This theory file contains 11 proofs by induction, many of which involve generalisation.
Previously, we reported low coincidence rates of \smartinduct{} for this file
and concluded that 
we could not achieve higher rates because of the domain-specific language we used to encode heuristics  \cite{smart_induction}:
this language, called \texttt{LiFtEr} \cite{lifter}, did not allow us to analyse definitions of relevant constants,
even though such definitions often carry the essential information to decide what variables to generalise.

As shown in Table \ref{table:nearest},
our evaluation confirmed low coincidence rates of \smartinduct{} for this file
but showed significantly higher rates of \smarterinduct{}.
That is,
\smarterinduct{} managed to predict experts' use of generalisation accurately
since 
\smarterinduct{} uses \selfie{} to analyse
the definitions of relevant constants as shown in Section \ref{sec:heuristic}.

\subsection{Return Rates for 5 Timeouts}\label{sec:eval_result_time}
\begin{table}
\begin{tabular}{c | r r r r r}
\hline
\noalign{\smallskip}
tool & 0.2s & 0.5s & 1.0s & 2.0s & 5.0s\\
\hline
\smarterinduct{} & 8.8\% & 24.7\% & 47.8\% & 69.8\% & 86.8\% \\ 
\texttt{smart}   & 0.0\% & 3.5\%  & 16.9\% & 38.3\% & 70.2\%\\ 
\hline
\end{tabular}
\caption{Return rates for five timeouts.}\label{table:semantic_induct_return}
\end{table}
\smarterinduct{} achieves the higher accuracy
by analysing not only the syntactic structures of inductive problems
but also the definitions of constants relevant to the problems.
Inevitably, this requires larger computational resources:
the \selfie{} interpreter has to 
examine not only the syntax tree representing proof goals
but also the syntax trees representing the definitions of relevant constants.
However,
thanks to
the syntax-directed candidate construction algorithm presented in Section \ref{sec:smart_construct}
and aggressive pruning strategy 
presented in Section \ref{sec:architecture},
\smarterinduct{} provides recommendations faster than \smartinduct{} does.

This performance improvement is presented in Table \ref{table:semantic_induct_return},
which shows how often \smarterinduct{} and \smartinduct{} return recommendations
within certain timeouts.
For example, the return rate of \smarterinduct{} is 8.8\% for 0.2 seconds.
This means that \smarterinduct{} returns recommendations 
for 8.8\% of proofs by induction within 0.2 seconds.
On the other hand, the return rate of \smartinduct{} is 0.0\% for 0.2 seconds.

Table \ref{table:semantic_induct_return} shows that
for all theory files
\smarterinduct{} produces more recommendations than \smartinduct{} does
for all timeouts specified in this evaluation,
proving the superiority of \smarterinduct{} over \smartinduct{} 
in terms of the execution time necessary to produce recommendations.
In fact, 
the median values of execution time for these 1,095 problems are
1.06 seconds for \smarterinduct{} and 
2.79 seconds for \smartinduct{}.
That is to say,
\smarterinduct{} achieved 62\% of reduction in the median value
of execution time.

\section{Related Work}
Boyer and Moore invented
the waterfall model \cite{waterfall}
for inductive theorem proving for a first-order logic on Common Lisp \cite{lisp}.
In the original waterfall model,
a prover tries to apply any of the following six techniques:
simplification, destructor elimination,
cross-fertilization, generalisation, elimination of
irrelevance, and induction
to emerging sub-goals until it solves all sub-goals.

The most well-known prover based on the waterfall model is ACL2 \cite{acl2}.
To decide how to apply induction,
ACL2 computes a score, called \textit{hitting ratio},
based on a fixed formula \cite{acl_book,induction}
to estimate how good each induction scheme is.
Instead of computing a hitting ratio,
we use \selfie{} to encode our induction heuristics as assertions.
While ACL2 produces many induction schemes and computes
the corresponding hitting ratios,
\smarterinduct{} produces a small number of 
promising sequences of induction terms and rules.

For Isabelle/HOL,
we developed a proof strategy language, PSL 
\cite{psl}.
PSL's interpreter discharges easy induction problems by conducting expensive proof searches,
and its extension to abductive reasoning tries to identify
auxiliary lemmas useful to prove inductive problems \cite{pgt}.
While our abductive reasoning mechanism took a top-down approach,
Johansson \etal{} took a bottom-up approach \cite{hipster} based on the idea of theory exploration.

There are ongoing attempts to extend saturation-based superposition provers with induction:
Cruanes presented an extension of typed superposition
that can perform structural induction \cite{zipperposition},
while Reger \etal{} incorporated lightweight automated induction \cite{vampire_induction}
to the Vampire prover \cite{vampire}
and Hajdú \etal{} extended it to cover induction with generalisation \cite{vampire_generalisation}.
A straightforward comparison to their approaches is difficult as
their provers are based on less expressive logics and different proof calculi.
However, we argue that one advantage of \smarterinduct{} over their approaches is that
\smarterinduct{} never introduces axioms that risk the consistency of Isabelle/HOL.
Furthermore, our evaluation consists of a wider range of problem domains 
written by experienced Isabelle users based on their diverse interests:
Hajdú's evaluation involved a number of inductive problems,
but the problem domains were limited to lists, natural numbers, and trees.

Similarly to \smarterinduct{},
TacticToe \cite{tactictoe,tactictoe_journal} for HOL4, Tactician \cite{tactician,tactictoe4coq} for Coq, and
PaMpeR \cite{pamper} for Isabelle are
meta-tactic tools seamlessly integrated in proof assistants' ecosystems; 
however, none of them logically analyse inductive problems
or predict arguments of \induct{} accurately.
Unlike these tools,
\smarterinduct{} presents accurate recommendations
without relying on statistical machine learning.

Despite the growing interest in deep learning for theorem proving,
\cite{holstep,deephol,CoqGym,enigma_anonymous,graph4HOL,topdown,taro_sekiyama,stateful_premise,deep_network_guided,isarstep},
we mindfully avoided deep learning
since we have only a limited number of inductive problems available. 
Instead of deep learning,
we used \selfie{}'s quantifiers to encode our heuristics in a domain-agnostic style.
To the best of our knowledge, 
no project based on deep learning has managed to predict arguments to \induct{} accurately.

\section{Conclusion}

We presented \smarterinduct{}, a recommendation tool for proof by induction.
\smarterinduct{} constructs candidate \texttt{induct} tactics for a given inductive problem
while avoiding combinatorial explosion, 
and it selects promising candidates by filtering out unpromising candidates
and scoring remaining ones.
To give scores to each remaining candidate,
we encoded 36 heuristics in \selfie{}
to decide on which terms and with which rules we should apply \induct{},
as well as 8 \selfie{} heuristic to decide which variables to generalise.

Our evaluation based on 1,095 inductive problems from 22 theory files showed that
compared to the existing tool, \smartinduct{},
\smarterinduct{} achieves a 90.0\% increase of the coincidence rate
from 20.1\% to 38.2\% for the most promising candidate,
while achieving a 62.0\% decrease of the median value of execution time.
In particular, \smarterinduct{} surpassed the accuracy of the existing tool
by a wide margin for inductive problems involving variable generalisation.

Currently, \smarterinduct{} uses manually specified weights for heuristics.
It remains as our future work to 
optimise such weights using evolutionary computation \cite{evolutionary_isabelle}
and to integrate \smarterinduct{} into a larger AI tool for Isabelle/HOL \cite{united}.

\section*{Acknowledgements}
We thank the anonymous reviewers for useful feedback both at TACAS2020 and IJCAI2021.
This work has been supported by the following grants:
\begin{itemize}
    \item the grant of Singapore NRF National Satellite of Excellence in Trustworthy Software Systems (NSoE-TSS),
    \item the European Regional Development Fund under the project AI \& Reasoning (reg.no.CZ.02.1.01/0.0/0.0/15\_
    003/0000466),
    \item the ERC starting grant no. 714034, and
    \item NII under NII-Internship Program 2019-2nd call.
\end{itemize}

\bibliographystyle{named}
\bibliography{ijcai21}

\begin{thebibliography}{}

\bibitem[\protect\citeauthoryear{Bansal \bgroup \em et al.\egroup
  }{2019}]{deephol}
Kshitij Bansal, Sarah Loos, Markus Rabe, Christian Szegedy, and Stewart Wilcox.
\newblock {HOL}ist: An environment for machine learning of higher order logic
  theorem proving.
\newblock In {\em Proceedings of the 36th International Conference on Machine
  Learning}, 2019.

\bibitem[\protect\citeauthoryear{Blaauwbroek \bgroup \em et al.\egroup
  }{2020a}]{tactictoe4coq}
Lasse Blaauwbroek, Josef Urban, and Herman Geuvers.
\newblock Tactic learning and proving for the {C}oq proof assistant.
\newblock In {\em {LPAR} 2020: 23rd International Conference on Logic for
  Programming, Artificial Intelligence and Reasoning, Alicante, Spain}, 2020.

\bibitem[\protect\citeauthoryear{Blaauwbroek \bgroup \em et al.\egroup
  }{2020b}]{tactician}
Lasse Blaauwbroek, Josef Urban, and Herman Geuvers.
\newblock The tactician - {A} seamless, interactive tactic learner and prover
  for {C}oq.
\newblock In {\em Intelligent Computer Mathematics - 13th International
  Conference, {CICM}, Bertinoro}, 2020.

\bibitem[\protect\citeauthoryear{Blanchette and Nipkow}{2010}]{nitpick}
Jasmin~Christian Blanchette and Tobias Nipkow.
\newblock Nitpick: {A} counterexample generator for higher-order logic based on
  a relational model finder.
\newblock In {\em Interactive Theorem Proving, First International Conference,
  {ITP} Edinburgh}, 2010.

\bibitem[\protect\citeauthoryear{Boyer and Moore}{1979}]{acl_book}
Robert~S. Boyer and J.~Strother Moore.
\newblock {\em A computational logic handbook}, volume~23 of {\em Perspectives
  in computing}.
\newblock Academic Press, 1979.

\bibitem[\protect\citeauthoryear{Bulwahn}{2012}]{quickcheck}
Lukas Bulwahn.
\newblock The new quickcheck for {I}sabelle - random, exhaustive and symbolic
  testing under one roof.
\newblock In {\em Certified Programs and Proofs - Second International
  Conference, {CPP}}, 2012.

\bibitem[\protect\citeauthoryear{Chvalovsk{\'{y}}}{2019}]{topdown}
Karel Chvalovsk{\'{y}}.
\newblock Top-down neural model for formulae.
\newblock In {\em 7th International Conference on Learning Representations,
  {ICLR} 2019, New Orleans, LA, USA}. OpenReview.net, 2019.

\bibitem[\protect\citeauthoryear{Cruanes}{2017}]{zipperposition}
Simon Cruanes.
\newblock Superposition with structural induction.
\newblock In Clare Dixon and Marcelo Finger, editors, {\em Frontiers of
  Combining Systems - 11th International Symposium, FroCoS 2017,
  Bras{\'{\i}}lia}, 2017.

\bibitem[\protect\citeauthoryear{Gauthier \bgroup \em et al.\egroup
  }{2017}]{tactictoe}
Thibault Gauthier, Cezary Kaliszyk, and Josef Urban.
\newblock Tactic{T}oe: Learning to reason with {HOL4} tactics.
\newblock In {\em LPAR-21, 21st International Conference on Logic for
  Programming, Artificial Intelligence and Reasoning, Maun}, 2017.

\bibitem[\protect\citeauthoryear{Gauthier \bgroup \em et al.\egroup
  }{2021}]{tactictoe_journal}
Thibault Gauthier, Cezary Kaliszyk, Josef Urban, Ramana Kumar, and Michael
  Norrish.
\newblock {T}actic{T}oe: Learning to prove with tactics.
\newblock {\em J. Autom. Reason.}, 65(2):257--286, 2021.

\bibitem[\protect\citeauthoryear{Gramlich}{2005}]{gramlich}
Bernhard Gramlich.
\newblock Strategic issues, problems and challenges in inductive theorem
  proving.
\newblock {\em Electr. Notes Theor. Comput. Sci.}, 125(2):5--43, 2005.

\bibitem[\protect\citeauthoryear{Hajd{\'{u}} \bgroup \em et al.\egroup
  }{2020}]{vampire_generalisation}
M{\'{a}}rton Hajd{\'{u}}, Petra Hozzov{\'{a}}, Laura Kov{\'{a}}cs, Johannes
  Schoisswohl, and Andrei Voronkov.
\newblock Induction with generalization in superposition reasoning.
\newblock In {\em Intelligent Computer Mathematics - 13th International
  Conference, {CICM}, Bertinoro}. Springer, 2020.

\bibitem[\protect\citeauthoryear{Harrison}{1996}]{hollight}
John Harrison.
\newblock {HOL} light: {A} tutorial introduction.
\newblock In {\em Formal Methods in Computer-Aided Design, First International
  Conference, {FMCAD}, Palo Alto}, 1996.

\bibitem[\protect\citeauthoryear{Jakubuv \bgroup \em et al.\egroup
  }{2020}]{enigma_anonymous}
Jan Jakubuv, Karel Chvalovsk{\'{y}}, Miroslav Ols{\'{a}}k, Bartosz Piotrowski,
  Martin Suda, and Josef Urban.
\newblock {ENIGMA} anonymous: Symbol-independent inference guiding machine
  (system description).
\newblock In {\em Automated Reasoning - 10th International Joint Conference,
  {IJCAR} Paris, France}. Springer, 2020.

\bibitem[\protect\citeauthoryear{Johansson \bgroup \em et al.\egroup
  }{2014}]{hipster}
Moa Johansson, Dan Ros{\'{e}}n, Nicholas Smallbone, and Koen Claessen.
\newblock Hipster: Integrating theory exploration in a proof assistant.
\newblock In {\em Intelligent Computer Mathematics - International Conference,
  {CICM} Coimbra, Portugal}, Lecture Notes in Computer Science. Springer, 2014.

\bibitem[\protect\citeauthoryear{Jr.}{1982}]{lisp}
Guy L.~Steele Jr.
\newblock An overview of common lisp.
\newblock In {\em Proceedings of the 1982 {ACM} Symposium on {LISP} and
  Functional Programming, {LFP} Pittsburgh}, 1982.

\bibitem[\protect\citeauthoryear{Kaliszyk \bgroup \em et al.\egroup
  }{2017}]{holstep}
Cezary Kaliszyk, Fran{\c{c}}ois Chollet, and Christian Szegedy.
\newblock Hol{S}tep: {A} machine learning dataset for higher-order logic
  theorem proving.
\newblock In {\em 5th International Conference on Learning Representations,
  {ICLR}, Toulon}. OpenReview.net, 2017.

\bibitem[\protect\citeauthoryear{Klein \bgroup \em et al.\egroup }{2004}]{AFP}
Gerwin Klein, Tobias Nipkow, Larry Paulson, and Rene Thiemann.
\newblock {\em The Archive of Formal Proofs}.
\newblock 2004.

\bibitem[\protect\citeauthoryear{Kov{\'{a}}cs and Voronkov}{2013}]{vampire}
Laura Kov{\'{a}}cs and Andrei Voronkov.
\newblock First-order theorem proving and vampire.
\newblock In {\em Computer Aided Verification - 25th International Conference,
  {CAV}}, 2013.

\bibitem[\protect\citeauthoryear{Li \bgroup \em et al.\egroup
  }{2021}]{isarstep}
Wenda Li, Lei Yu, Yuhuai Wu, and Lawrence~C. Paulson.
\newblock Isar{S}tep: a benchmark for high-level mathematical reasoning.
\newblock In {\em International Conference on Learning Representations}, 2021.

\bibitem[\protect\citeauthoryear{Loos \bgroup \em et al.\egroup
  }{2017}]{deep_network_guided}
Sarah~M. Loos, Geoffrey Irving, Christian Szegedy, and Cezary Kaliszyk.
\newblock Deep network guided proof search.
\newblock In {\em LPAR-21, 21st International Conference on Logic for
  Programming, Artificial Intelligence and Reasoning, Maun, Botswana}, EPiC
  Series in Computing. EasyChair, 2017.

\bibitem[\protect\citeauthoryear{Moore and Wirth}{2013}]{induction}
J~Strother Moore and Claus{-}Peter Wirth.
\newblock Automation of mathematical induction as part of the history of logic.
\newblock {\em CoRR}, abs/1309.6226, 2013.

\bibitem[\protect\citeauthoryear{Moore}{1973}]{waterfall}
J.~Strother Moore.
\newblock {\em Computational logic : structure sharing and proof of program
  properties}.
\newblock PhD thesis, University of Edinburgh, {UK}, 1973.

\bibitem[\protect\citeauthoryear{Moore}{1998}]{acl2}
J.~Strother Moore.
\newblock Symbolic simulation: An {ACL2} approach.
\newblock In {\em Formal Methods in Computer-Aided Design, Second International
  Conference, {FMCAD} '98, Palo Alto}, 1998.

\bibitem[\protect\citeauthoryear{Nagashima and He}{2018}]{pamper}
Yutaka Nagashima and Yilun He.
\newblock Pa{M}pe{R}: proof method recommendation system for {Isabelle}/{HOL}.
\newblock In {\em Proceedings of the 33rd {ACM/IEEE} International Conference
  on Automated Software Engineering, {ASE}, Montpellier}, 2018.

\bibitem[\protect\citeauthoryear{Nagashima and Kumar}{2017}]{psl}
Yutaka Nagashima and Ramana Kumar.
\newblock A proof strategy language and proof script generation for
  {I}sabelle/{HOL}.
\newblock In {\em 26th International Conference on Automated Deduction,
  Gothenburg}, 2017.

\bibitem[\protect\citeauthoryear{Nagashima and Parsert}{2018}]{pgt}
Yutaka Nagashima and Julian Parsert.
\newblock Goal-oriented conjecturing for {I}sabelle/{HOL}.
\newblock In {\em Intelligent Computer Mathematics - 11th International
  Conference, {CICM}, Hagenberg}, 2018.

\bibitem[\protect\citeauthoryear{Nagashima}{2019a}]{lifter}
Yutaka Nagashima.
\newblock Li{F}t{E}r: Language to encode induction heuristics for
  {Isabelle}/{HOL}.
\newblock In {\em Programming Languages and Systems - 17th Asian Symposium,
  {APLAS}, Nusa Dua, Bali, Indonesia}, 2019.

\bibitem[\protect\citeauthoryear{Nagashima}{2019b}]{evolutionary_isabelle}
Yutaka Nagashima.
\newblock Towards evolutionary theorem proving for {I}sabelle/{HOL}.
\newblock In {\em Proceedings of the Genetic and Evolutionary Computation
  Conference Companion, {GECCO}, Prague, Czech Republic}, 2019.

\bibitem[\protect\citeauthoryear{Nagashima}{2020a}]{selfie}
Yutaka Nagashima.
\newblock {SeLFiE}: Modular semantic reasoning for induction in
  {I}sabelle/{HOL}.
\newblock {\em CoRR}, abs/2010.10296, 2020.

\bibitem[\protect\citeauthoryear{Nagashima}{2020b}]{smart_induction}
Yutaka Nagashima.
\newblock Smart induction for {I}sabelle/{HOL} (tool paper).
\newblock In {\em Proceedings of the 20th Conference on Formal Methods in
  Computer-Aided Design – FMCAD}, 2020.

\bibitem[\protect\citeauthoryear{Nagashima}{2020c}]{united}
Yutaka Nagashima.
\newblock Towards united reasoning for automatic induction in {Isabelle}/{HOL}.
\newblock In {\em The Japanese Society for Artificial Intelligence 34th Annual
  Conference, {JSAI}, online}, volume
  \url{https://doi.org/10.11517/pjsai.JSAI2020.0_3G1ES103}, 2020.

\bibitem[\protect\citeauthoryear{Nagashima}{2021}]{appendix}
Yutaka Nagashima.
\newblock Appendix to ``faster smarter proof by induction in
  {I}sabelle/{HOL}''.
\newblock {\em Zenodo, \url{https://doi.org/10.5281/zenodo.4743750}}, 2021.

\bibitem[\protect\citeauthoryear{Nipkow \bgroup \em et al.\egroup
  }{2002}]{isabelle}
Tobias Nipkow, Lawrence~C. Paulson, and Markus Wenzel.
\newblock {\em Isabelle/HOL - a proof assistant for higher-order logic}.
\newblock Springer, 2002.

\bibitem[\protect\citeauthoryear{Paliwal \bgroup \em et al.\egroup
  }{2020}]{graph4HOL}
Aditya Paliwal, Sarah~M. Loos, Markus~N. Rabe, Kshitij Bansal, and Christian
  Szegedy.
\newblock Graph representations for higher-order logic and theorem proving.
\newblock In {\em The Thirty-Fourth {AAAI} Conference on Artificial
  Intelligence, {AAAI}}. {AAAI} Press, 2020.

\bibitem[\protect\citeauthoryear{Piotrowski and Urban}{2020}]{stateful_premise}
Bartosz Piotrowski and Josef Urban.
\newblock Stateful premise selection by recurrent neural networks.
\newblock In {\em {LPAR} 2020: 23rd International Conference on Logic for
  Programming, Artificial Intelligence and Reasoning, Alicante, Spain}, EPiC
  Series in Computing, 2020.

\bibitem[\protect\citeauthoryear{Reger and Voronkov}{2019}]{vampire_induction}
Giles Reger and Andrei Voronkov.
\newblock Induction in saturation-based proof search.
\newblock In {\em {CADE} - 27th International Conference on Automated
  Deduction, Natal}, 2019.

\bibitem[\protect\citeauthoryear{Sekiyama and Suenaga}{2018}]{taro_sekiyama}
Taro Sekiyama and Kohei Suenaga.
\newblock Automated proof synthesis for the minimal propositional logic with
  deep neural networks.
\newblock In {\em Programming Languages and Systems - 16th Asian Symposium,
  {APLAS}, Wellington, New Zealand}. Springer, 2018.

\bibitem[\protect\citeauthoryear{Slind and Norrish}{2008}]{hol4}
Konrad Slind and Michael Norrish.
\newblock A brief overview of {HOL4}.
\newblock In {\em Theorem Proving in Higher Order Logics, 21st International
  Conference, TPHOLs Montreal}, 2008.

\bibitem[\protect\citeauthoryear{{T}he {C}oq~development team}{2021}]{coq}
{T}he {C}oq~development team.
\newblock {T}he {C}oq proof assistant, \url{https://coq.inria.fr}, 2021.

\bibitem[\protect\citeauthoryear{Wenzel}{2002}]{isar}
Markus Wenzel.
\newblock {\em {I}sabelle, {I}sar - a versatile environment for human readable
  formal proof documents}.
\newblock PhD thesis, Technical University Munich, Germany, 2002.

\bibitem[\protect\citeauthoryear{Wenzel}{2012}]{jedit}
Makarius Wenzel.
\newblock Isabelle/j{E}dit - {A} prover {IDE} within the {PIDE} framework.
\newblock In {\em Intelligent Computer Mathematics - 11th International
  Conference}, 2012.

\bibitem[\protect\citeauthoryear{Yang and Deng}{2019}]{CoqGym}
Kaiyu Yang and Jia Deng.
\newblock Learning to prove theorems via interacting with proof assistants.
\newblock In {\em Proceedings of the 36th International Conference on Machine
  Learning}, 2019.

\end{thebibliography}

\end{document}